# Optimizing Traffic Signal Control using High-Dimensional State Representation and Efficient Deep Reinforcement Learning

Lawrence Francis[1], Blessed Guda[2], Ahmed Biyabani[3]
[123]*Carnegie Mellon University Africa, Kigali, Rwanda*
Email: [1]*lfrancis@andrew.cmu.edu, 2 blessedg@andrew.cmu.edu,* [3] *ab3x@andrew.cmu.edu*

**Abstract:** In reinforcement learning-based (RL-based) traffic signal control (TSC), decisions on the signal timing are made based on the available information on vehicles at a road intersection. This forms the state representation for the RL environment which can either be high-dimensional containing several variables or a low-dimensional vector. Current studies suggest that using high dimensional state representations does not lead to improved performance on TSC. However, we argue, with experimental results, that the use of high dimensional state representations can, in fact, lead to improved TSC performance with improvements up to 17.9% of the average waiting time. This high-dimensional representation is obtainable using the cost-effective vehicle-to-infrastructure (V2I) communication, encouraging its adoption for TSC. Additionally, given the large size of the state, we identified the need to have computational efficient models and explored model compression via pruning.

## 1. Introduction

Road traffic congestion results in increased fuel consumption and emissions, and accounts for 75% of the 24% global $CO_2$ emissions from transportation [1], [2], [3]. Road intersections are usually equipped with some sensing capabilities for vehicle detection, enabling the design of algorithms which take the current state of the intersection into consideration to produce an optimal signal timing plan. A common approach is the use of Reinforcement Learning (RL) which has shown promising results compared to other approaches [4]. The RL agent learns by simulating interactions with traffic environments, where they are rewarded for reducing delay or improving throughput, allowing systems to self-optimize and adapt effectively to traffic patterns[5], [6]. However, the performance of the RL agent is highly dependent on how detailed it can sense the environment and measure its performance in that environment. The typical sensors used in these RL solutions use loop detectors that capture only a low state representation data like the number of vehicles per lane or even cameras that can capture individual vehicle positions and speeds[7] . However, even more variables can be captured using Vehicle-to-Infrastructure(V2I) communication where vehicles send data to traffic signals using V2I technology [7].

With V2I communication, traffic signals can have access to a rich amount of information on incoming vehicles, including position, speed, waiting time and delay. This gives rise to a high-dimensional state representation to be used by an RL agent for traffic signal control. Figure 1 shows a TSC system using V2I communication to send data from On-Board Units (OBU) to a Roadside Unit (RSU). The RSU processes and transfers the received data to the controller which controls the traffic lights.

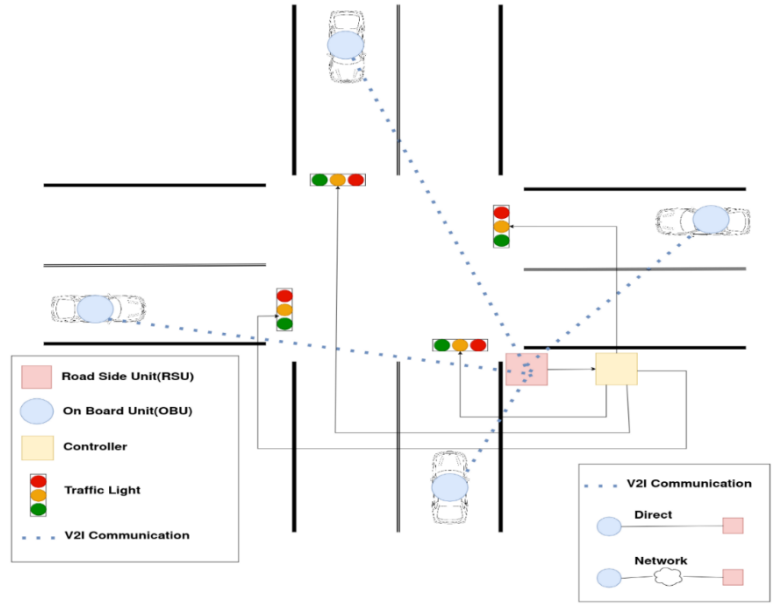

*Figure. 1. A Traffic signal control system using V2I communication to detect and obtain vehicle data. The OBUs sends vehicle data (position, speed, waiting times, etc) to the RSU and the RSU processes and transfers this data to the controller which controls the traffic lights at the intersection.*

While loop detectors and cameras are commonly used as sensors, they are usually quite expensive to install and require frequent maintenance. Also, the number of loop detectors or cameras to be installed is usually proportional to the number of lanes at the intersection. On the other hand, using V2I communication, with technologies like DSRC and Cellular-V2X, only requires the installation of a single RSU per intersection and an OBU should come pre-installed on vehicles [8], [9], [10] . This makes the use of V2I communication more cost-efficient than the commonly used detectors today[11]. The cost benefits would help in enhancing the scalability of Intelligent Transportation Systems (ITS) in resource-constrained settings like Africa [12]

An RL-based traffic signal controller relies on an environment design of the intersection. This environment provides a state representation, which encompasses the various factors and conditions present, and a reward function, guiding the RL agent's decision-making for traffic signal timing. The dimensionality of the state representation is determined by the sensing capabilities around the traffic signal environment. Low-dimensional vector representations can be used to represent the lane occupancy information obtained from loop detectors. An example is the number of vehicles on each lane, with the state represented by a vector whose length is equal to the number of incoming lanes and each element is the number of detected vehicles for that lane.

Several studies have favoured the use of low-dimensional vector representations of environment state, arguing that the use of additional information in environment state does not lead to additional performance improvements[11], [13], [14]. This has influenced the need to settle for a simpler low-dimensional state representation. However, we argue that the rich vehicle information contained in the high-dimensional state representation obtainable from V2I communication can lead to improved performance on TSC. The increased computational demand associated with processing high-dimensional data using large deep learning models can lead to higher energy consumption and $CO_2$ emissions [15]. This creates a double bind for using high-dimensional states. To address this, we propose the use of a parameter-efficient Deep RL model, incorporating pruning techniques to reduce the computational load. This approach aims to strike a balance between leveraging the rich data available from V2I communication and maintaining an environmentally friendly and efficient system by minimizing the computational footprint.

In this research, we choose to use a Deep Q-Network (DQN) RL model since it has shown promising performance in TSC. Training the DQN requires choosing some hyperparameters like learning rate and discount factor to obtain optimal results. We show that, with careful hyperparameter tuning, we can achieve better performance with high-dimensional state representations. The improvements brought about by having a rich amount of vehicle information in the state representation suggest that V2I communication, which is cost effective, can be used as an alternative to traditional detectors for traffic signal control. We also highlight the benefits and downsides of using these state representations.

# 1. Background and Related Works

Several works have explored the use of reinforcement learning for traffic signal control and the effect of the traffic signal environment state representation on the performance of reinforcement learning algorithms. Intellilight [5] used a DQN-based agent to conduct extensive experiments using real world traffic datasets while providing several interpretations of the learned policy. An investigation into the challenge of defining the state and rewards for traffic signal control was carried out in [11] where it was observed that queue length as a reward function and simple state representation like the number of vehicles in each lane gives better. This idea of using simpler state representation is further propagated into future works as is seen in PressLight [16] and MPLight [17]which use the number of vehicles in each lane as state and intersection pressure as reward function.

The use of V2I communication for vehicle detection in traffic signal control was explored in a research conducted by [18], where RL-based algorithms were used for traffic signal control under partial vehicle detection. This work highlighted the cost effectiveness of using V2I communication as opposed to using loop detectors and cameras for sensing at signalized intersections. A recent work highlights the importance of the rich information provided by V2I communication for traffic signal control as opposed to other detection methods[7].

Most works that investigate environment state representations all conclude, from experimentation, that using simpler state representations is better because of the reduced complexity and similar performance to complex image-like representations [11], [13], [14]. However, we identify the benefit of high-dimensional state representations and the rich vehicle information that is obtained using V2I communication for an improved RL-based traffic signal control.

Pruning techniques involving removing model parameters that contribute less to overall model performance have been explored for deep learning models showing that the number of model parameters can be reduced with no significant decrease in the model accuracy [19]. Pruning value-based deep reinforcement learning methods for certain levels of sparsity can lead to better performance than an original dense model [20]. For the real-world application and implementation of RL-based traffic signal control, the storage and computational requirements of the model to be deployed needs to be taken into consideration. For RL-based traffic signal control, experimental results suggests that an efficient RL-based solution incorporating pruning to attain a sparse model with fewer parameters has been shown to deliver optimal performance on traffic signal control tasks [21].

*1.1 Reinforcement Learning*

The goal of RL is for an agent to maximize a cumulative reward starting from the initial state till the terminal state [22]. Starting from an initial state, an agent takes an action and is transitioned into the next state and given an immediate reward. This process continues until a terminal state is encountered. The performance of the agent is then evaluated using the sum of rewards received throughout the episode. This is formalised as a Markov Decision Process (MDP) defined by the tuple $(S, A, r, T, \gamma)$ where S is the set of states, $A$ is the set of actions, $r$ is a reward function, $T$ is a state transition probability function, and $\gamma$ is a discount factor. The discount factor is introduced in the learning process of a policy to favour present rewards over future rewards. The policy is thus learned using a discounted cumulative reward. The environment that the agent learns from is designed to incorporate these elements of the MDP.

In a traffic signal control environment, the state is represented by modelling the information that the traffic signal sensors can capture about the vehicles on all incoming lanes to the intersection. The choice of these state representations is heavily influenced by the sensors. The action space contains all non-conflicting phases. A phase is a set of non-conflicting traffic movements allowed to occur simultaneously. The action space is thus discrete and expressed as $A = \{0, 1, 2 \ldots, P - 1\}$ where $P$ is the number of phases for the traffic signal. The reward is usually a value that should be maximized for optimal performance, and for TSC, we maximize the negative queue length. A value function is the expected cumulative reward obtainable from any state and a popular value-based algorithm is the Deep Q-Network (DQN) [23].

## 2. Methodology

We aim to optimize the flow of traffic at signalized intersections, utilizing the rich information about incoming vehicles at an intersection. We make use of a high-dimensional image-like state representation for this vehicular information and show the performance increase brought about by having more information in the state representation, as opposed to using simpler vector state representations. We also introduce a simple model compression scheme for reducing model size and efficiency.

*2.1 – Environment state representations*

The environment state representation for a traffic signal environment is designed using the vehicle information on each incoming lane of the intersection and accounts for the sensing capabilities of the traffic signal sensor. The state can be represented using the aggregated vehicle information for each lane or direct vehicle information. Using the aggregated vehicle information for each lane gives rise to vector state representation where the state is a vector, and each element represents a lane measure [24]. The state can also be represented using direct vehicle information stored in a multidimensional array. The choice of vehicle information and lane measure to include in the state representation depends on the choice of sensor and how practical it is to obtain such information using the sensor in the real world.

**Vector state representations:** For the vector state representation, we compute lane measures as aggregated vehicle information for each lane. Possible lane measures include the number of vehicles and average queue length obtainable with loop detectors, and average waiting time and average delay that can be obtained using V2I technology. For a traffic intersection with $L$ incoming lanes, we represent the vector state representation as $S_L \in R^L$. $S_L = [s_1, s_2, \ldots, s_L]$ where $s_i$ is the lane measure for lane $i$ for $i = 1, \ldots, L$. For any traffic signal environment, the vector state representation can either contain the number of vehicles, average waiting time or average queue length.

**High-dimensional state representation:** The state representation can also account for individual vehicles at the intersection, storing their information in a multidimensional array. Given a vehicle detection radius, that indicates how close to the traffic signal a vehicle must be to be detected, a multidimensional array can be constructed to contain the information obtainable from all detected vehicles. This creates a rich high-dimensional state representation for the traffic signal environment. For a traffic intersection with $L$ incoming lanes, and a detection radius $R$ from the traffic signal, we represent the high-dimensional state representation as $H_L \in R^{R \times L \times m}$ where $m$ is the number of variables being considered for the vehicle information. Possible variables to consider are position, speed, and waiting time of the vehicles, in which case $m = 3$. This results in an "Image-like" [14] state representation

These state representations are described in Figure 2. We considered 3 environment state representations: **Number of vehicles per lane**, **Average waiting time per lane**, and **Image-like state representation**.

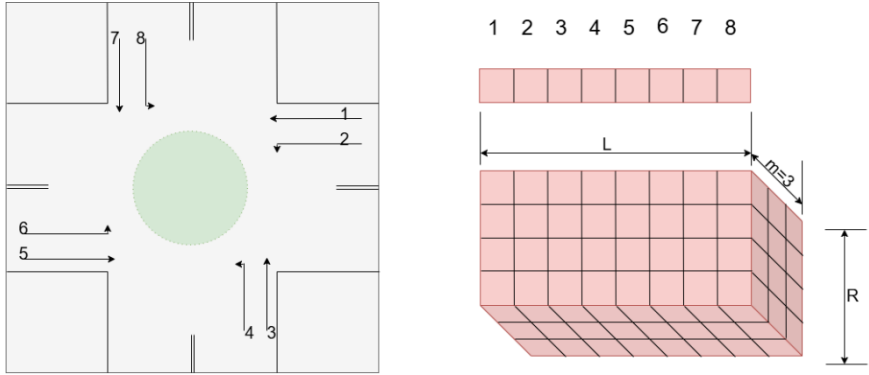

*Figure 2: Vector state representation and Image-like representation for an intersection with 8 incoming lanes. The vector state representation shown in the top right is a vector having 8 elements and each element corresponds to an aggregated lane measure. The image-like state representation is depicted at the bottom right. It can store individual vehicle information within a given radius, R, from the intersection, for each lane. We have chosen to capture only 3 variables for the vehicle information, so m=3.*

## 2.2 –Traffic Signal Controller

We use a DQN agent as the traffic signal controller. The agent has a model (called the Q-Network) a replay buffer for storing experiences from the environment. The Q-network computes the value of taking an action in a given state. Figure 2 shows the training setup of the DQN agent. The agent interacts with the environment to generate transitions which are tuples of state, action, reward, and next state. These transitions are stored in the replay buffer and sampled intermittently to be used for model update. To reduce the size of the Q-Network, it undergoes a sparse pruning process where a certain percentage of the parameters are pruned.

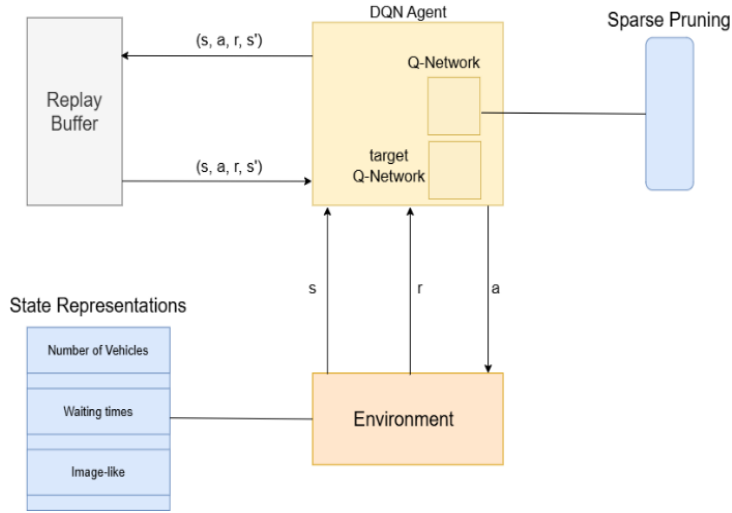

*Figure 2: This figure shows the interaction between the DQN agent and the environment. The environment has a choice of state representations to use. The agent interacts with the environment by choosing an action (a) given a state (s), the environment responds with the next state (s') and reward (r). The agent uses a replay buffer to store transitions it encounters in the environment and uses these transitions to update the Q-Network. The Q-Network undergoes sparse pruning for size reduction.*

*2.3 Model Compression by Pruning*

To alleviate the computational issues brought about by having larger models, we employed a magnitude-based model pruning strategy where model parameters that do not contribute so much to the model performance are pruned. Using the magnitude-based pruning approach [19], we compressed the model to about half the original size.

## 3. Experiments And Results

To support our argument about the improved performance attainable with high-dimensional state representations, we design experiments involving evaluating the performance of an RL agent in optimizing the traffic flow at an intersection as measured by some evaluation metrics when different state representations are used. Given that value-based agents like DQN perform quite well for traffic signal control [14], we use DQN as the RL agent. We evaluate the performance of three state representations; 1.) Number of vehicles per lane, 2.) Average waiting time per lane 3.) Image-like state representation [14]. The actual experiment runs are preceded by a hyperparameter search sweep that determines the best hyperparameters to use when training the DQN agent for each environment state representation.

Our experiments were conducted using SUMO (Simulation of Urban Mobility) [25] simulation software and two road networks from the "Cologne" road network provided in the RESCO benchmark [26], as shown in Figure 3. Each simulation for traffic in the road networks lasts for 3600 seconds (1 hour). Our environment-agent interface was designed using the "sumo-rl" library [27].

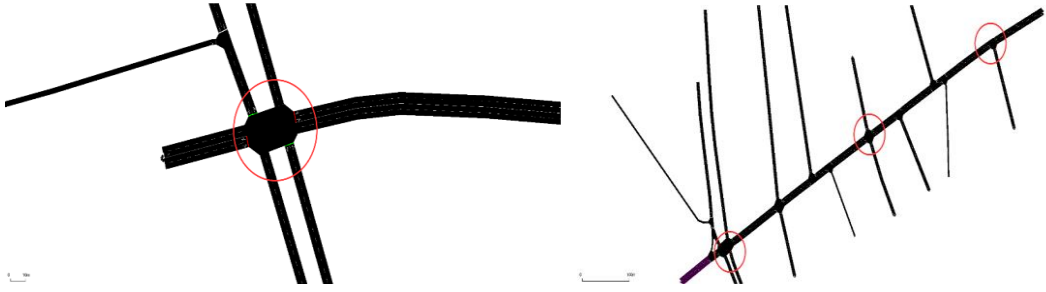

*Figure 3: The "Cologne" road networks used for evaluation. The network at the left contains only one signalized intersection and would be referred to a cologne1. The network at the right contains 3 signalized intersections. The signalized intersections are highlighted using red circles.*

*4.1 Evaluation metrics*

The following evaluation metrics were used to evaluate the agent's performance, all time are in seconds.
- **Average travel time (ATT)**: The average travel time is the average time take for a vehicle to arrive at its destination having travelled from its starting position. It is computed by adding up the travel times of all vehicles in a road network and dividing by the total number of vehicles.
- **Average waiting time (AWT)**: The waiting time is the average time that a vehicle spent idle at the intersection while waiting for the right of way.
- **Average delay (AD)**: The delay is the difference between actual travel time of a vehicle and the expected travel time. The average is obtained as the sum of all vehicle delays divided by the total number of vehicles.
- **Average queue length (AQL)**: The queue length at an intersection is the number of vehicles on queue at the intersection, and the average queue length is the sum of the number of vehicles on queue on each incoming lane of the intersection divided by the number lanes.

*4.2 Hyperparameter search*

For the hyperparameter search, we implemented a grid search over the three hyperparameters: discount factor, learning rate, target update frequency. We searched over 3 possible values of each hyperparameter. The values searched over for each hyperparameter are as follows:
- Discount factor: 0.9, 0.95, and 0.99
- Learning rate: 0.1, 0.01, 0.001
- Target update frequency: 5, 10, 15

Using 3 possible settings over 3 hyperparameters results in 27 possible hyperparameter combinations. For each hyperparameter combination, we perform 3 training runs for 70 episodes. The mean of the resulting metrics for the 3 runs are computed and we select the combination with the minimum score as measured by the evaluation metrics.

*4.3 DQN Agent performance*

Using a DQN agent, and the obtained hyperparameter for each state representation, we trained the DQN agent for each state representation for 200 episodes over 5 runs with different random seeds. We report the best agent performance for the evaluation metrics. The best scores on the evaluation metric for the "Cologne1" road network with a single intersection is reported in Table 1 while that of the "Cologne3" road network with 3 intersections is reported in Table 2. Using the image-like representation resulted in an improvement in performance for all metrics.

Table 1: Agent performance for the road network with a single intersection (Cologne1)

| State Representations | Metric | | | |
|---|---|---|---|---|
| | ATT | AD | AWT | AQL |
| Number of Vehicles | 40.36 | 30.15 | 11.24 | 6.95 |
| Average waiting time | 43.33 | 32.94 | 13.14 | 7.20 |
| Image-like | **39.64** | **29.43** | **10.79** | **6.65** |

Table 2: Agent performance for the road network with 3 intersections (Cologne3)

| State Representations | Metric | | | |
|---|---|---|---|---|
| | ATT | AD | AWT | AQL |
| Number of Vehicles | 56.57 | 43.68 | 12.04 | 3.65 |
| Average waiting time | 57.00 | 43.93 | 13.28 | 4.09 |
| Image-like | **56.49** | **43.62** | **11.64** | **3.48** |

*4.4 Pruning performance*

For the agent using the image-like state representation, we examine the effect of model pruning on the performance, considering several levels of sparsity up to 50%. Figure 4 shows the performance at several sparsity levels, with 15% sparsity having the best performance.

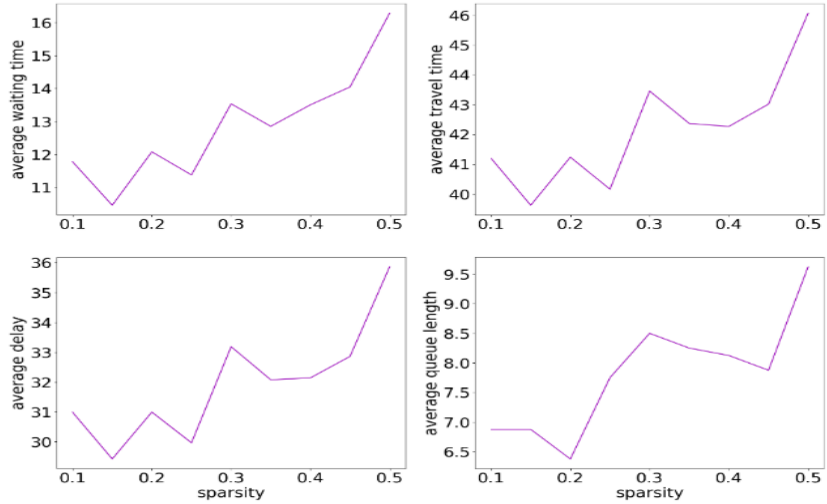

Figure 4: Agent's performance for several levels of sparsity up to 50% (0.5). The best performing sparsity level is 15% (0.15).

## 4. Discussions

Our experimental results in Table 1 and Table 2 show that both for Cologne1 and Cologne3, the image-like representation lead to improvements across all metrics compared to the two vector state representations. In the Cologne1 scenario, it reduced the average travel time (ATT) by 1.8% compared to the number-of-vehicles representation and 8.5% compared to the average-waiting-time representation. Similar improvements were observed in average delay (AD), average waiting time (AWT), and average queue length (AQL), with reductions ranging from 2.4% (AD for number-of-vehicles) to 17.9% (AWT for average-waiting-time representation).

Even In the more complex three-intersection scenario (Cologne3), the image-like representation maintained better performance across all metrics. The best performance improvement was 14.91% for AQL when using the image like representation over the average waiting time. We also notice a 3.3% improvement in the AWT over the best performing vector state representation. This suggests that the benefits of rich state representation scale to multi-intersection scenarios.

The model compression via pruning results in Figure 4 show the trade-off between model size and performance. The optimal pruning level of 15% sparsity suggests that some model parameters are indeed redundant, and their removal improves performance. This finding aligns with recent research showing that sparse networks can sometimes generalize better than their dense counterparts [20], [21]. However, the sharp performance degradation beyond 15% sparsity indicates a clear limit to how much the model can be compressed while maintaining performance. In our experiment, we have only performed model compression after the training process, this is termed post training pruning. Better compression performance can be attained by performing in-training pruning, where the model size of being reduced during training by removing some model parameters. This is an area to explore in future works using high-dimensional state representations.

Given the improved performance obtained using a high-dimensional state representation as evident in the observed experimental results, we can infer that the use of high-dimensional state representations can lead to a more optimal traffic flow at signalized intersections. An optimal traffic flow reduces congestion, leading to environmental, social and economic benefits like less fuel consumption, less $CO_2$ emissions, and reduced stress for commuters. Also, we identify that having more information in the state representation requires the use of a sensor capable of capturing this information at signalized intersections, and the conventional loop detectors are not suitable for this. With V2I communication at intersections, several vehicle information (like speed, position, waiting time, destination, etc.) can be captured, since this can be sent directly from the vehicles. With this improved performance we reported, the adoption of V2I technology at signalized intersections should be encouraged and given the cost benefits of using V2I, we can have more signalized intersections.

## 5. Conclusions

This research demonstrates that high-dimensional state representations, enabled by V2I data can improve the performance of deep reinforcement learning-based traffic signal control systems. Our results show consistent improvements across multiple evaluation metrics for both single and multi-intersection scenarios, challenging the conventional intuition that simpler state representations are sufficient for optimal performance.

However, the complexity of high-dimensional state representation poses challenges in real-world deployment, especially in terms of computational demands and the environmental impact of training such large models. To address this, we employed model compression through pruning, which maintains high performance while making the model more efficient. Future work can focus on refining efficient training techniques to further minimize the environmental impact of model development, advancing scalable, sustainable traffic management solutions for smart cities.